\documentclass[usenatbib,onecolumn]{mnrasarx}
\usepackage{graphicx}	
\usepackage{amsmath}	
\usepackage{amssymb}	

\title[Microwave pulsations from a twisted loop]{Pulsations of microwave emission from a solar flare in a twisted loop caused by intrinsic MHD oscillations}

\author[C. Smith et al.]{
Christopher Smith,$^{1}$
M. Gordovskyy,$^{1}$
and P.K. Browning$^{1}$
\\
$^{1}$Department of Physics and Astronomy, University of Manchester, Oxford Road, Manchester M13~9PL, UK}

\date{In original form November 2021}

\pubyear{2021}

\begin{document}
\label{firstpage}
\pagerange{\pageref{firstpage}--\pageref{lastpage}}
\maketitle

\begin{abstract}
We present results revealing microwave pulsations produced in a model of a flaring twisted solar coronal loop, without any external oscillatory driver. Two types of oscillations are identified: slowly-decaying oscillations with a period of about 70--75~s and amplitude of about 5--10\% seen  in loops both with and without energetic electrons, and oscillations with period of about 40~s and  amplitude of a few tens of percent observed only in  loops with energetic electrons for about 100~s after onset of fast energy release. We interpret the longer-period oscillations as the result of a standing kink mode modulating the average magnetic field strength in the loop, whilst the short-period intermittent oscillations associated with energetic electrons are likely to be produced by fast variations of the electric field which produces  energetic electrons in this scenario. The slowly-decaying oscillations can explain the quasi-periodic pulsations often observed in the flaring corona.
\end{abstract}


\section{Introduction}
\label{s-intro}

Waves and oscillations are ubiquitous in the solar corona \cite[e.g.][]{mona12,nako20}. During flares and other active events, they can be observed in nearly all parts of the electromagnetic spectrum, from decametric radio to hard X-ray, with periods ranging from tens of milliseconds to tens of minutes \cite[e.g.][]{nake99,nake04,zist09,kume16}. Understanding the origin and properties  of these waves and oscillations is essential for understanding  energy release and transport in solar flares. Furthermore, waves can serve as  unique diagnostic tools for physical conditions in the solar corona \citep{mona12,chpe15}.

One type of oscillation observed in both solar and stellar flares is the so-called quasi-periodic pulsations (QPPs): intensity oscillations with typical periods in the range 1--100s and amplitudes 1--10\% observed in extreme UV, hard and soft X-ray, and microwave (MW) ranges. The exact nature of these oscillations is not fully understood and is a matter of heated debate \cite[e.g.][]{dooe16,mcce18,zime21}. Although the observed periods clearly indicate that QPPs are the result of some magnetohydrodynamic (MHD) process, it is not clear whether the observed oscillations are a direct effect of magnetic field oscillations in the corona or a proxy. For instance, the presence of QPPs in hard X-ray emission may indicate that MHD oscillations modulate particle acceleration in the corona, which, in turn, results in an oscillatory pattern in the intensity of bremsstrahlung produced by non-thermal electrons \cite[e.g.][]{mcce18,care19,haye19}. Another key question is whether the observed QPPs are due to auto-oscillations of the coronal magnetic field structure, or instead  forced or triggered externally, for example by an incoming MHD wave.

In this study, we  address these questions through forward modelling, combining resistive-MHD and radiative transfer simulations, and using a semi-empirical model of the non-thermal electron  population motivated by the results of MHD-test-particle simulations. We model a confined flare in a twisted coronal loop, in which energy is released by magnetic reconnection. We investigate the time-domain properties of the MW emission produced by this  flaring loop, showing that pulsations similar in many ways to observed QPPs can be produced without any external wave-like modulation.

\begin{figure*}
\centering{\includegraphics[width=1.0\textwidth]{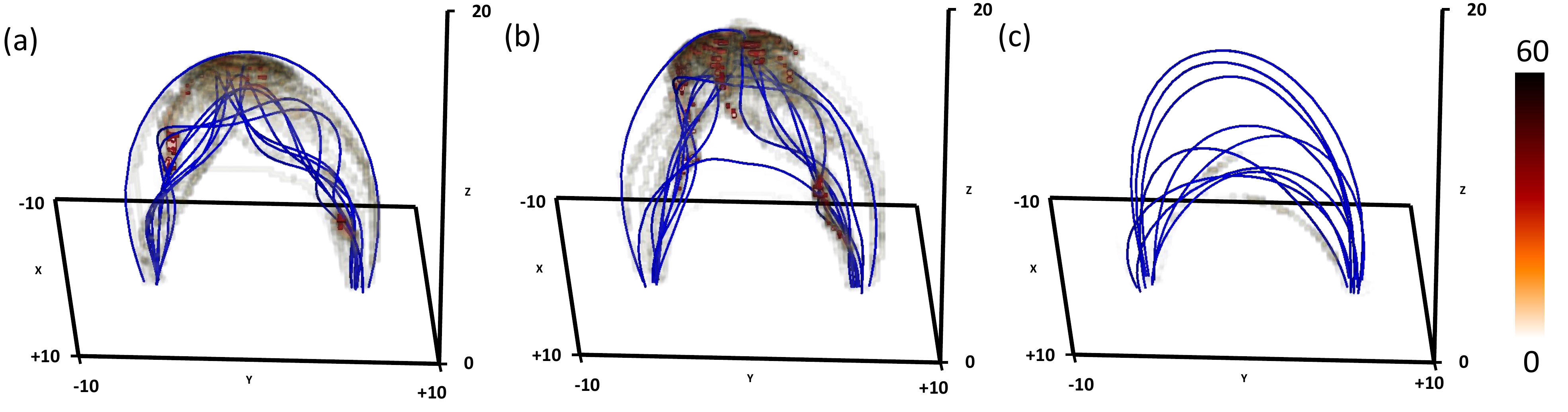}}
\caption{Evolution of the magnetic and electric field in the considered model. Blue lines denote selected magnetic field lines, while the semi-transparent volume plot show distribution of the parallel electric field in the domain. The corresponding colour scale shows values in V~m$^{-1}$ units. Panels (a), (b) and (c) correspond to t=2~s, 80~s and 200~s, respectively, after the onset of the  kink instability.}
\label{f-model}
\end{figure*}

\section{The model}
\label{s-model}

The model of a confined solar flare in a twisted magnetic loop is based on the scenario developed by \cite{gobr11,gore14} by combining 3D MHD and test-particle simulations \cite[MHDTP][]{gore10}. In this model magnetic energy release is triggered by the onset of ideal kink instability \citep{broe08}. The instability leads to magnetic reconnection,  heating the plasma \citep{hooe09} and accelerating electrons and ions \citep{gore14}. The properties of thermal and non-thermal emissions in this model have been studied by \citet{gore14,gore17} and \citet{pine16}.

We consider a  twisted coronal loop with  length of about 120~Mm, cross-section varying from about 4~Mm at foot-points to about 14~Mm at the summit and the magnetic field varying from about 500~G at the base to about 50~G at the top. The loop is embedded within a  gravitationally-stratified atmosphere with the temperature of 10$^4$K in a 2~Mm-wide layer just above the lower boundary of the domain (representing the chromosphere), and 0.9~MK everywhere else. The evolution of magnetic field and thermal plasma is calculated using 3D resistive-MHD LARE3D code \citep{arbe01} (Figure~1). At the onset of  magnetic reconnection, the magnetic field is twisted with the twist angle of about 4$\pi$ (Figure~1a). Eventually, the loop kinks, resulting in localised enhancements of the current density, which, in turn triggers anomalous resistivity, leading to  magnetic reconnection and energy release on a timescale of about 100~s (Figure~1b). The twist is reduced, lowering the magnetic energy,  and the loop also becomes wider, due to reconnection between the loop field and the ambient field (Figure~1c). The high current densities and enhanced resistivity in some locations also  result in high electric fields, which accelerates particles. 

\citet{gore17} investigated the MW emission produced in this configuration and showed that the MW polarisation properties can be used as a diagnostic of the magnetic twist. This effect has subsequently been observed by \citet{shae18}. In this study we use a  similar approach, but focus on the time domain. We utilise a radiative transfer code built on the code by \citet{flku10}.  This  code can simulate generation and transfer of free-free and GS emission from fast electrons in the MW spectral range, assuming that at every point within the model domain the energy distribution of electrons is as a sum of a Maxwellian distribution with  density $n_t$ and temperature $T_t$, and a power-law non-thermal electron distribution with density $n_b$, power-law index $\gamma$, lower and upper energy cut-offs $E_{1}$ and $E_{2}$, respectively, analytical pitch-angle distribution.

The MW radiative transfer is modelled on a sparse numerical grid. The original 256$\times$256$\times$512 grid from the MHD simulations is degraded to 64$\times$64$\times$64 with the grid step of 1.25~Mm in all directions. This substantially speeds-up calculations, while the resulting resolution ($\sim$1~arcsec) is still better than most solar MW observations. 

Using test-particles from MHDTP simulations directly would result in very high noise due to undersampling: it is impossible to evaluate the parameters of energetic electrons for each cell, given there may be  only about 5 test-electrons per cell. Thus instead,  energetic electron parameters are calculated analytically in a way which mimics the behaviour of energetic electrons in the MHDTP models \citep{gore14,pine16,gore17}. 

The spatial distribution of energetic electrons is determined by the amplitude and structure of the electric field component parallel to the magnetic field, which is calculated using Ohm's law:
$E_{||}=\vec{B}\cdot\left(\eta\vec{j}-\left(\vec{v}\times \vec{B}\right)\right)/B$.
The number of energetic electrons in a given location $\vec{r}$ of the spatially-degraded grid is defined as 
$N(\vec{r})=0.01\left(\tilde{E}_{||}/E_{tr}\right)\tilde{n_t}$, where $\tilde{n_t}$ and $\tilde{E}_{||}$ are the values of the thermal plasma density and parallel electric field averaged along the magnetic field. The above formula means that the fraction of electrons, that get  accelerated is proportional to the local parallel electric field. The constant  $E_{tr}$ is set to be 70 V m$^{-1} $, this scaling factor is chosen to achieve a total fraction of non-thermal electrons typical of solar flares.

Averaging along the magnetic field reflects the fact that energetic electrons move very quickly, and, hence, redistribute along field lines on timescales by factor of about $v_A/c$ shorter than the timescale of processes in the MHD model (where  $v_A$ and $c$ are the Alfven velocity and the speed of light, respectively). The process of deriving values of plasma density $\tilde{n_t}$, temperature $\tilde{T_t}$, and parallel electric field $\tilde{E}_{||}$ has been first implemented in \citet{wate20}.

The energy and pitch-angle distribution of high-energy electrons is defined as 
$f~\sim~E^{-\gamma}\exp\left(-(\mu-1)/0.04\right)$, limited by the lower and upper energy cut-offs, $E_1$ and $E_2$, respectively.
Here $E$ is the electron kinetic energy, $\mu = v_{||}/v$ is the pitch-angle cosine, where $v_{||}$ is the component of electron velocity parallel to magnetic field, $\gamma$ is power-law spectral index, $\Theta$ is the Heaviside (or step-like) function.  The lower and upper energy cut-offs are taken to be $E_1=10$~keV and $E_2=10$~MeV, respectively, at all times throughout the  domain. The results in Section~3 are produced using electron energy distribution with $\gamma=3$.

The spatial structure of degraded parallel electric field $\tilde{E_{||}}$ is shown in Figure~1.  Note the fragmented distribution of the parallel electric field, corresponding to the distribution of current sheets described above.

The observed radiation intensity (Stokes I) is forward-modelled assuming that the loop is viewed from the side, i.e. the MW intensity maps are produced for $y-z$ plane by integrating radiative transfer equations in $x$-direction.  

\section{Microwave intensity oscillations}
\label{s-results}

\begin{figure*}
\centering{\includegraphics[width=1.0\textwidth]{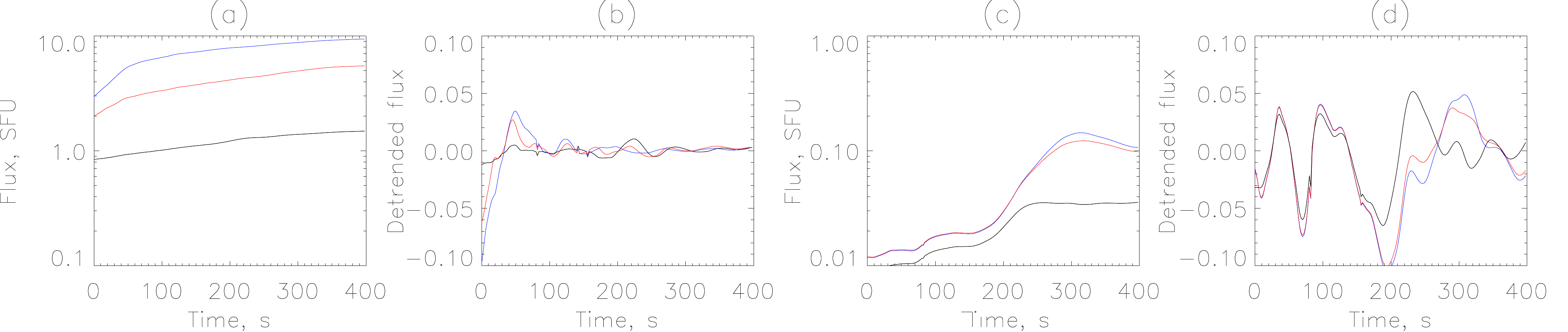}}
\caption{Light-curves (panels a and c) and de-trended light-curves (panels b and d) for the model with no particle acceleration. Horizontal axes show time after the onset of magnetic reconnection. Panels (a-b) and (c-d) correspond to the emission from the whole domains and loop-tops, respectively. Black, red and blue curves correspond to MW frequency of 5, 15 and 30~GHz. De-trended values (panels b and d) are normalised by the trend values.}
\label{f-lctherm}
\end{figure*}
\begin{figure*}
\centering{\includegraphics[width=1.0\textwidth]{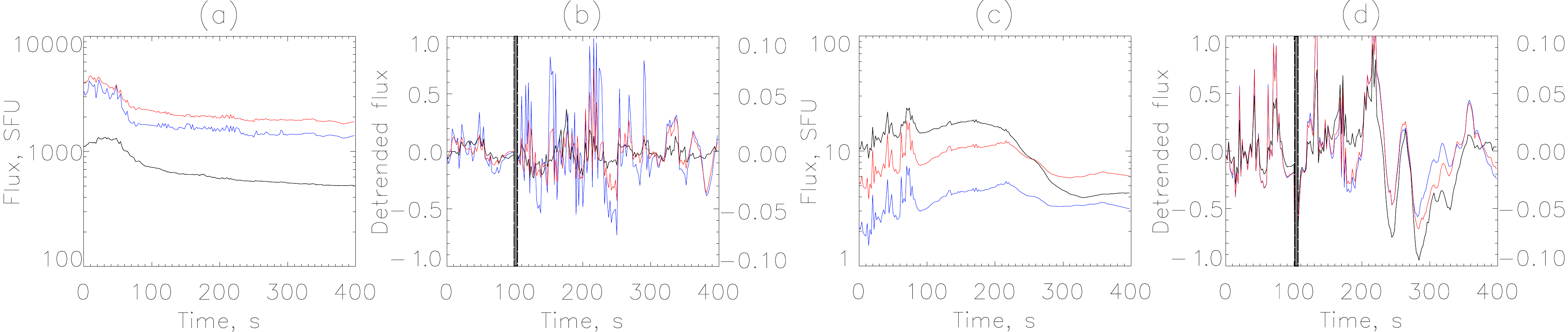}}
\caption{The same as in Figure~\ref{f-lctherm}, but for the model with non-thermal particles. There is a break in panels (b) and (d): de-trended values corresponding to times between t=0 and t=100~s have different axis scaling.}
\label{f-lcnonth}
\end{figure*}
\begin{figure*}
\centering{\includegraphics[width=1.0\textwidth]{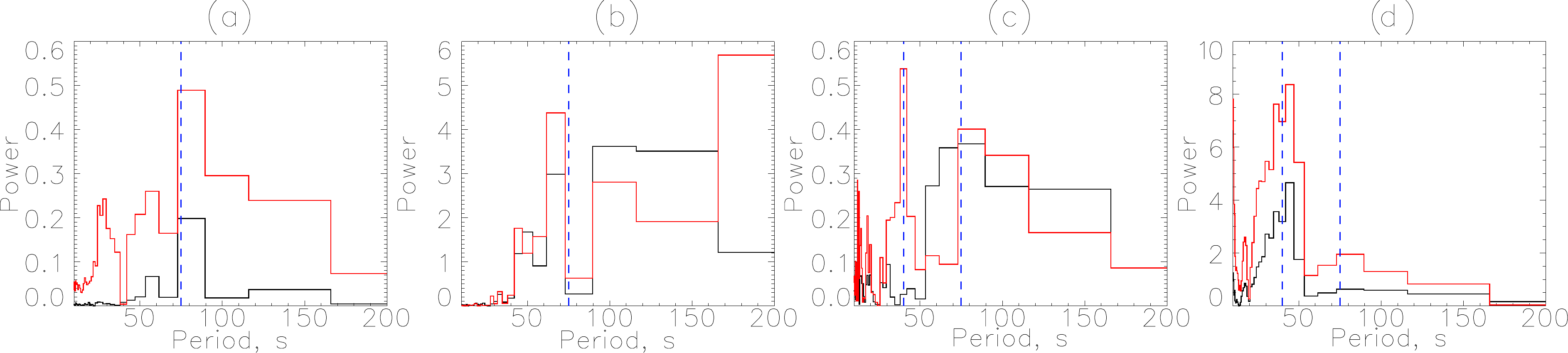}}
\caption{Power spectra of MW intensity oscillations. Panels (a-b) correspond to the  fully thermal model, panels (c-d) correspond to the model with non-thermal particles as well as thermal plasma. Panels (a) and (c) correspond to the MW emission integrated over the  whole domain, panels (b) and (d) correspond to the loop-top emission. The black and red lines are for MW frequency of 5 and 15~GHz, respectively. The blue dashed lines denote periods of 40 and 75~s.}
\label{f-pspectra}
\end{figure*}
\begin{figure}
\centering{\includegraphics[width=0.27\textwidth]{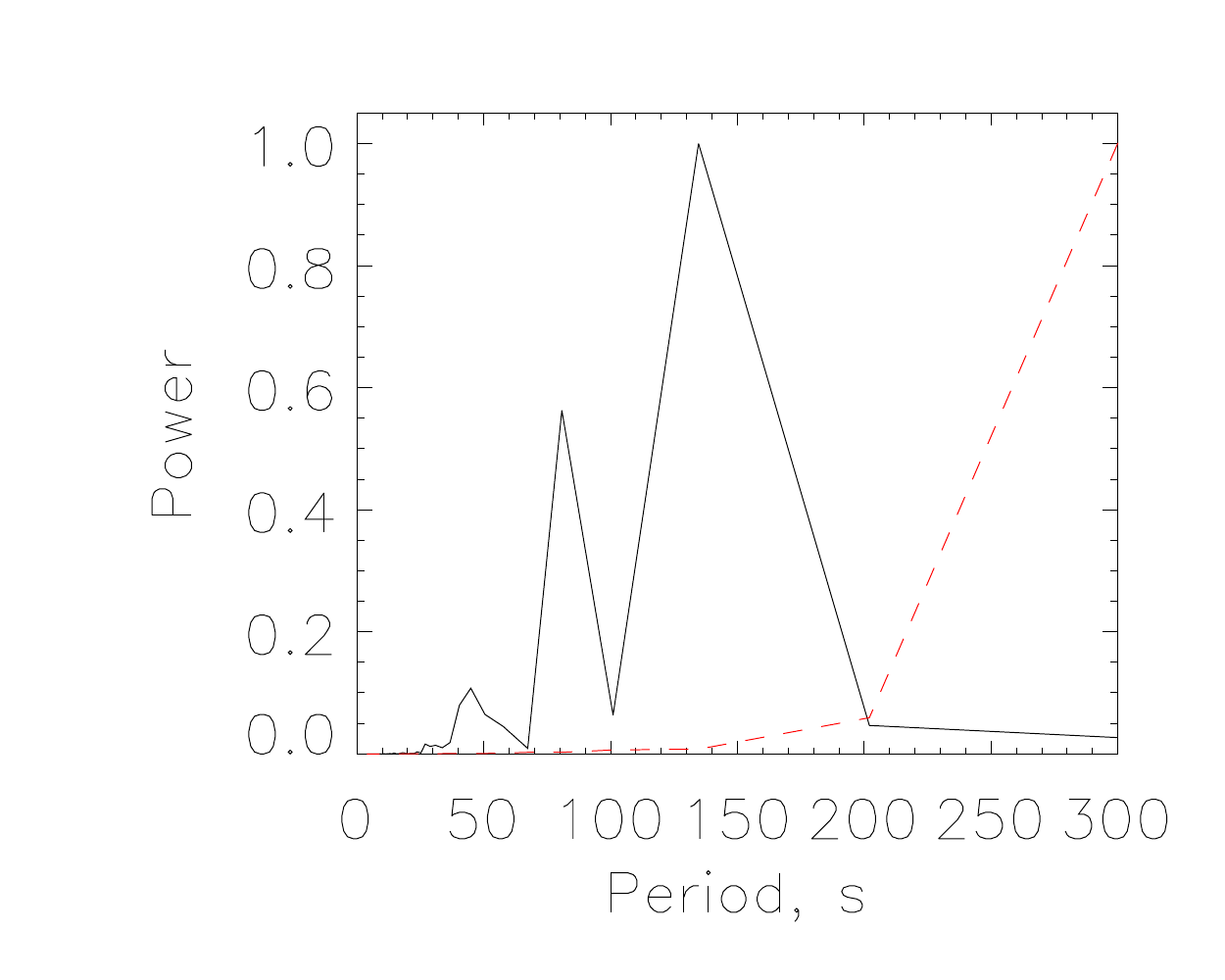}}
\caption{Power spectra of the LOS velocity variation (solid black line) and of average density variation (dashed red line) in the mid-plane of the loop.}
\label{f-kink}
\end{figure}

MW emission from the two models is calculated for frequency range 1--100GHz. In both models emission appears to be dominated by gyrosynchrotron. In the purely thermal model the spectral slope is positive in the whole considered spectral range. In the model with energetic particles, the spectral slop is positive under the turning frequency of about 30--35~GHz.

The study focuses on the lower frequencies, 5--30~GHz. This is because several instruments extensively used to study time-domain properties of MW emission in solar flares, such as Nobeyama Radio Heliograph, observed in this frequency range.

We analyse emission produced by the whole loop and, separately, by the loop-top region. The latter is bounded by heights of 32 and 48~Mm above the lower boundary in the  vertical ($z$) direction, and by -8 and +8~Mm with respect  to the  loop summit location in the horizontal ($y$) direction. MW light curves are shown in the left panels of Figure~2, for purely thermal plasma, and in Figure~3, for the case with non-thermal electrons in addition to the thermal plasma.

In order to detect any oscillatory signal, a de-trending procedure is carried out in a way similar to that used in observational data processing \cite[e.g.][]{broe19}.
The trend is determined by smoothing the original light-curves with the 100~s sliding window, with the edge values truncated. After the trend is subtracted, the remaining function is normalised by the trend function, giving the relative amplitude of the oscillatory component. 

The characteristics of the MW emission from the purely thermal model are shown in the Figure~2. The MW intensities from this model slowly increase with time. This reflects the fact that the plasma heating, which is distributed within the loop due to plasma flows and thermal conduction, happens on relatively long timescales of about 200--300~s, in contrast with the Ohmic heating itself, which happens approximately within the first 100~s after onset of magnetic reconnection \citep{pine16}. The increase of intensity from the loop-top is delayed compared to the intensity from the loop as a whole. This is because  the energy release in this model is not concentrated near the loop top, it is distributed throughout  the loop. In fact, because the magnetic field and current density are higher closer to the foot-points (because of the magnetic field convergence), most of the direct Ohmic heating occurs in the loop legs. 

The de-trended lightcurves for the purely thermal case (Figure~2 b, d)  show distinct  oscillations with a typical period in the range 50--100~s. Oscillations of the emission from the whole loop initially have an amplitude of about 5\%, but quickly decay, showing an amplitude of 1\% after 300~s at all three considered frequencies (5, 15 and 30~GHz). However, oscillations of the emission from at the loop-top show no signs of decay: the amplitude remains about 5\% at all three frequencies. This difference between the loop-top and the whole loop is likely to be due to the fact that the oscillations of intensity produced in the loop legs is smeared by fast changes in the energy release and heating.

Oscillations of the emission coming from the loop top at the three different frequencies appear to be synchronised, unlike the intensity of the loop as a whole. This indicates that the radiation from the loop-top is mostly optically-thin, while the emission generated in the loop legs is optically-thick \cite[see e.g.][]{gore17}. As the result, the loop-top emission 'samples' the same physical volume along the line-of-sight, while the emission from the loop legs at different frequencies 'samples' different layers of the loop. 

The emission from the model containing non-thermal electrons is significantly different (Figure~3). Firstly, MW intensity peaks at about 100~s after the onset of the reconnection. After about 200~s it monotonically decreases. This reflects the fact that most of the emission is produced by energetic electrons, while particle acceleration is significant during the magnetic reconnection stage (about first 100~s). During the first $\sim$100~s, MW emission shows strong variations at all frequencies with the amplitudes of up to 50\% with the characteristic period of $\sim$20--30~s. The oscillations of MW intensity from the whole loop look almost like white noise, without any periodicity. At the same time, pulsations of MW emission from the loop-top seem to have a period of about 15--35~s. At the later stages, 100--400~s after onset of the reconnection, the loop lightcurves show oscillations with  amplitude of about 5\% and  periods in the range 50--100~s, similar to the 'thermal' loop. However, in the model with non-thermal particles, these relatively low frequency oscillations are smeared with the higher frequency variations, yielding non-stationary pattern of pulsations, similar to those observed in some solar and stellar flares \citep{nake19}.

In order to understand time domain properties of the obtained lightcurves, we perform their Fourier analysis at each MW frequencies. The power spectra of the oscillations as functions of period are shown in Figure~4. As expected, the emission from the loop-top region in the purely thermal model reveals a relatively narrow, intense peak corresponding to period of around 60--70~s. There is a significant signal at longer periods ($>$100--200~s), reflecting slower variations. In principle, these slow variations are the 'residuals' left by the de-trending procedure. It is impossible to clearly separate the trend and the oscillations because (a) there is no clear difference in periods and (b) the period of oscillations is in  order-of-magnitude comparable to the modelling time interval, i.e. edge effects, inevitably, affect the power spectrum. 

The power spectra corresponding to the whole domain also show a clear peak, although it is slightly shifted to longer periods, 70--80~s. Interestingly, both radio-frequencies shown in this graph, 5 and 15~GHz, have main peaks appearing at the same periods. This means that, although oscillations at different frequencies are not synchronised, this only affects the phase, but not the period.

The main difference between the power spectra corresponding to models with and without energetic particles is that the former model reveals additional peaks at shorter periods, about 40s. This peak corresponds to the oscillations, appearing during the first 100s. It is an interesting finding: although the corresponding interval of the light-curve seems to be quite noisy, the peaks are quite narrow.

The question regarding the nature of MHD oscillations resulting in the observed quasi-periodic MW pulsations remains open. The MW pulsations with the period of about 70--80~s are consistent with a standing MHD mode with the wavelength comparable to the loop length: the ratio of the loop length ($\sim$120~Mm) to the average Alfven speed ($\sim$1.4$\times$10$^6$~m~s$^{-1}$) is about 90~s. The MW intensity can be modulated either by variation of the line-of-sight (LOS) component of the magnetic field due to the kink mode, or by variation of the number of energetic electrons due to sausage mode. The fact that we see clear pulsations in the nearly-optically-thin emission, at 30GHz, from the loop-top in the case with energetic particles, indicates variation of the LOS component of magnetic field, meaning that MW pulsations are likely caused by the kink mode. Furthermore, analysis of the velocity field in the model reveals the presence of kink mode with a period close to the period of MW pulsations. Figure~5 compares power spectrum of the average LOS velocity in the mid-plane of the loop, which is a proxy for the kink mode, with the  power spectrum of average density, which is a proxy of the sausage mode. The former shows peaks at 130--140~s and around 75~s; both periods can result in the MW pulsations with the period of 70--75~s. At the same time, the density power spectrum has nearly power-law shape with no peaks. Hence, the longer period MW pulsations are most likely caused by the large-scale kink.

The nature of higher frequency oscillations in the model with energetic electrons is less clear. The sausage mode is more effective in modulating the current density and, hence, the parallel electric field, which, in turn, would modulate the number of energetic particles.  The fact that its period is approximately half of the longer period oscillations may indicate that it is a higher harmonic global MHD mode modulating the average current density and parallel electric field.  

\section{Summary}
\label{s-summary}

Our analysis of  synthetic MW emission produced by thermal and non-thermal plasma in 3D magnetohydrodynamic simulations  of  reconnecting twisted coronal loops, finds MW intensity  oscillations  in the frequency range 5--30~GHz,  without any external driver.

A simulated  coronal loop with the length of about 120~Mm and average magnetic field of about 70~G exhibits  oscillations with periods of about 70--80s and an amplitude of about 5\% during about 300--400s. Since these oscillations are observed in models both with and without energetic electrons, we interpret them to be due to a standing global MHD mode modulating the average magnetic field in the loop, which, in turn, modulates the GS emission produced by hot thermal plasma.

In addition to these oscillations, the loop with energetic electrons demonstrates an interval of  higher-frequency oscillations with the period of about 40~s.
We interpret these oscillations as the result of a higher-order MHD mode modulating the average electric field in the loop, and, hence, modulating the  number  of energetic electrons, which produce GS emission during the energy release stage (about 100s).

Our key finding here is the existence of these oscillations, and this opens up the several significant  areas for future work, including more detailed studies of observables. We believe that the kink mode is more likely to be responsible for the longer period MW pulsations, while fast oscillations observed immediately after onset of reconnection are caused either by higher-order kink or sausage mode. However, this issue will be addressed in more detail  in a forthcoming study. 

The obtained characteristics of the MW emission oscillations are very similar to those in solar flare  quasi-periodic pulsations observed in the MW spectral range \cite[e.g.][]{zime21}. Therefore, we conclude that auto-oscillations of a flaring loop can produce observed QPPs in the absence of  any external oscillatory modulation.

\section*{Acknowledgment}
MG and PKB were supported by the Science and Technology Facilities Council (STFC, UK), grant ST/T00035X/1.

\section*{Data availability}
Lare3D code is available at warwick.ac.uk/fac/sci/physics/research/cfsa/people/tda/larexd/. The MW radiative transfer code is included into GX simulator framework, part of the SolarSoft, mssl.ucl.ac.uk/surf/sswdoc/solarsoft/ssw\_install\_howto.html. The data underlying this article will be shared on reasonable request to the corresponding author.

\bibliographystyle{mnras}
\bibliography{bibs}

\end{document}